\documentclass[prd,twocolumn,nofootinbib,preprintnumbers,showpacs]{revtex4-1}

\usepackage{amssymb,amsmath,amsfonts,amsthm,latexsym,nicefrac,bm,bbm,esint,braket,url,epsfig,graphicx,epstopdf} 

\begin{document}

\preprint{DESY-14-150}

\title{{\Large Power Spectrum of Inflationary Attractors}}

\author{Benedict J.\ Broy${}^1$, Diederik Roest${}^2$ and Alexander Westphal${}^1$}

\affiliation{{}$^1$Deutsches Elektronen-Synchrotron DESY, Theory Group, \\22603 Hamburg, Germany}

\affiliation{{}$^2$Van Swinderen Institute, University of Groningen, Nijenborgh 4, 9747 AG Groningen, The Netherlands}

\begin{abstract}
Inflationary attractors predict the spectral index and tensor-to-scalar ratio to take specific values that are consistent with Planck.  An example is the universal attractor for models with a generalised non-minimal coupling, leading to Starobinsky inflation. In this paper we demonstrate that it also predicts a specific relation between the amplitude of the power spectrum and the number of e-folds. The length and height of the inflationary plateau are related via the non-minimal coupling: in a wide variety of examples, the observed power normalisation leads to at least 55 flat e-foldings. Prior to this phase, the inflationary predictions vary and can account for the observational indications of power loss at large angular scales.
\end{abstract}


\maketitle

\textbf{Introduction.} Recent years have seen staggering progress in the field of CMB observation. High accuracy measurements have transformed the once speculative field of cosmology into a precision science. Combined data from WMAP and Planck \cite{1212.5225,1303.5082}  provides strong evidence for a very early phase of cosmological inflation. The CMB gives constraints on inflation that are by and large consistent with the picture of simple single-field slow-roll inflation. Yet the very same observations contain hints \cite{1303.5075} that the power of the CMB temperature spectrum may be suppressed by 5\%-10\% at large angular scales ($\ell\lesssim 40$) compared to a spectrum with $n_s$\thinspace=\thinspace\thinspace$0.960$ and no running. Similarly, the COBE results \cite{9601067} already contain evidence towards the same effect. Cosmic variance \cite{0303072} limits any measurement of the $c_\ell$ to $\Delta c_\ell \sim (2\ell+1)^{-1/2}$.  At low-$\ell$, the Planck temperature TT data already reaches this limit. At smaller scales, $\Delta c_{\ell}$ is not yet reached experimentally everywhere, and adding future data may still lead to slight variations of the value of $n_s$. Adding future polarization data will provide additional independent data at low-$\ell$ in form of the TE and EE correlations. Moreover, future large-scale structure surveys and 21-cm tomography may provide even more modes at low-$\ell$ due to an increased sample volume compared to the CMB alone \cite{1309.4060}. Thus the significance of the observed power loss may still change considerably in the future \cite{1309.4060}. Finally, if the B-mode polarization detected by BICEP2 \cite{BICEP2} corresponds to a primordial tensor mode signal from inflation with $r\sim 0.1$, this would roughly double the amount of power suppression hinted for by the CMB data at low-$\ell$ \cite{1404.2278}. 

If one assumes this apparent power loss to be a real effect, the question arises whether we can find natural mechanisms that induce the required running of the spectral index $n_s$ without invoking heavy fine-tuning. To achieve this, we turn our attention to a general class of inflation models referred to as universal attractors \cite{1310.3950}. We will  investigate whether generic corrections to an underlying functional relation inherent to those models give rise to sufficient running of $n_s$. In this class of models, the running of $n_s$ and the overall normalisation of the inflaton potential are linked via the requirement of having an inflationary plateau of certain length. We will find that, requiring $\sim 55$ flat e-folds of inflation, a natural class of corrections \emph{predicts} the normalisation of the CMB spectrum as well as the percentage of power loss to the order of magnitude in accordance with observation. Furthermore, we will outline the consequences of this setup for Higgs and eternal inflation.\\ 
\\
\textbf{Power Loss at low-$\ell$.} We will start by reviewing how a steepening in the scalar field potential suppresses power in the temperature power spectrum at low-$\ell$, if placed at the onset of observable e-folds \cite{0303636,1211.1707,1309.3413, 1309.3412,1309.4060, whipped, 1404.2278,1407.1048} (various other sources of power suppression are studied in e.g.\ \cite{0303636,1407.1048}). At large angular scales, the temperature power spectrum of the primordial curvature perturbation is given by \cite{09075424}
\begin{equation}\label{spectrum}
l(l+1)C^{TT}_l\propto\Delta^2_s(k) \propto \left(k/k_*\right)^{n_s-1},
\end{equation}
where $n_s=1+2\eta_V-6\epsilon_V$ is the spectral index, $\eta_V, \epsilon_V$ are the potential slow-roll parameters and $k_*$ is a pivot scale. Power loss arises when the spectral index increases with decreasing $k$. To obtain power loss within the first observable e-folds, we require $n_s$ to fall sufficiently fast. We thus parametrise the scalar field equation in terms of the number of e-folds $N_e=\ln(a/a_{end})$ with\footnote{We approximate inflationary spacetime as de Sitter space, thus $a\propto e^{Ht}$ with $H$ being the Hubble parameter during inflation. $N_e$ is negative throughout inflation and becomes zero at the end of inflation. As a shorthand, $n_s(62)$ means $n_s$ at $N_e=-62$.} $a=a_{end}e^{Ht}$, where the precise value of $a_{end}$ depends on the details of reheating and shall not concern us any further. We get
\begin{equation}\label{sce}
\chi''+3\left(1-\chi'^2/6 \right)\left(\chi'+\partial_\chi \ln V \right)=0,
\end{equation}
with $()'=d/dN_e$ and $\chi$ being the inflaton.
In slow roll, $\chi''\ll\chi'$ and thus $\chi'\approx -\partial_\chi\ln V$. Hence \eqref{sce} may be solved numerically to give $\chi(N_e)$. We can then evaluate the slow-roll parameters (setting $8\pi G=M_{Pl}=1$)
\begin{align}
\epsilon_V=\left(\partial_\chi \ln V \right)^2/2,\quad \eta_V=\left(\partial^2V/\partial\chi^2\right)/V,
\end{align}
on the numerical solution to investigate whether $n_s$ falls off sufficiently fast. At last, to identify $N_e$ with the wave number $k$, we recall that a mode $k$ exits the horizon when $k=a_kH_k$, where $H_k$ denotes the inverse event horizon during inflation and $a_k$ is the size of the scale factor at horizon exit. Thus
\begin{equation}
k=a_k H_k=a_{end} e^{N_e}H_k,
\end{equation}
where $N_e<0$. Rearranging, we find
\begin{equation}
N_e(k)=\log\left(\frac{k}{a_0H_0}\right)-\log\left(a_{end}\frac{H_k}{H_0}\right) ,
\end{equation}
in terms of  today's Hubble parameter. The second term on the right hand side is $\sim 62$, the exact value again depending on the details of reheating and the inflationary energy scale. From the above we find that the scale $k_*=0.05$ $Mpc^{-1}$ left the horizon at $N_e\sim -55$. Having a relation between wave number $k$ and number of e-folds $N_e$, we may investigate \eqref{spectrum} with $n_s$ being dependent on $k$ through $N_e$. To obtain the percentage of suppression $\%(N_e)$, we can then compare $\Delta_s^2(k)$ at the onset of observable e-folds to a spectrum with no running of $n_s$.
Finally, we emphasize that the slow-roll conditions, i.e.\ $|\epsilon_V, \eta_v|< 1$, hold throughout the inflationary trajectory, including the observable interval with power loss. \\
\\
\textbf{Universal Attractor.} The class of inflationary models referred to as the universal attractor \cite{1310.3950} is based on a specific non-minimal coupling to gravity, which we may view as a  generalisation of Higgs inflation \cite{Salopek, Bezrukov08} to arbitrary potentials. Consider the Jordan frame Lagrangian of a scalar field with non-minimal coupling,
\begin{equation}\label{JordanL}
\frac{\mathcal L_J}{\sqrt{-g_J}}=\frac{1}{2}\Omega(\phi)R_J-\frac{1}{2}g_J^{\mu\nu}\partial_\mu\phi\partial_\nu\phi-V_J(\phi),
\end{equation}
where 
\begin{align} 
\Omega(\phi)=1+\xi f(\phi),\quad V_J(\phi)=\lambda^2 f(\phi)^2.\label{1}
\end{align}
Note the functional relation between the scalar potential and the non-minimal coupling: both are expressed in terms of a single function $f(\phi)$ that we will take to be vanishing at the origin. This will be referred to as the attractor relation. Furthermore, the  strength of the non-minimal coupling is set by the parameter $\xi$ that is crucial for the attractor behaviour. The scalar potential has an overall parameter $\lambda^2$ that plays no role in the inflationary predictions of the universal attractor; for consistency with the original work, we keep the parameter but will assume it takes a natural value of order one.

One can transform to the Einstein frame  with the standard Einstein-Hilbert term $R/2$ via
\begin{equation}
g^E_{\mu\nu}=\Omega(\phi)g^J_{\mu\nu},
\end{equation}
where the superscripts denote Einstein and Jordan frame respectively. The Lagrangian \eqref{JordanL} then becomes
\begin{equation}\label{EinsteinL}
\frac{\mathcal L_E}{\sqrt{-g_E}}=\frac{1}{2}R_E-\frac{1}{2}\left[\frac{1}{\Omega}\left(\partial\phi \right)^2+\frac{3}{2}\left(\partial \ln \Omega \right)^2 \right] -\frac{V_J}{\Omega^2}. 
\end{equation}
At sufficiently large $\xi$ (for which in many cases $\xi >1$ suffices), the second term of the kinetic energy dominates, which we refer to as the strong coupling regime. Hence it is natural to consider the following redefinition
\begin{equation}
\frac{1}{2}\left(\partial \chi \right)^2=\frac{3}{4}\left(\partial \ln\Omega \right)^2,
\end{equation}
in order to introduce a canonically normalised kinetic term. This leads to
\begin{equation}\label{omega}
\Omega(\chi)=e^{\sqrt{2\over 3}\chi}.
\end{equation}
Rewriting $f$ in terms of $\Omega$ and evaluating the expression for the potential of \eqref{EinsteinL} with \eqref{omega} gives
\begin{equation}\label{staro}
V_E=\lambda^2\thinspace\xi^{-2}\left(1-e^{-\sqrt{\frac{2}{3}}\chi} \right)^2,
\end{equation}
where we identify $\chi$ as the inflaton. Thus any scalar field with potential $V_J$ and frame function $\Omega-1\sim \sqrt{V_J}$ in the Jordan frame is, at strong coupling, conformally equivalent to $R^2$-inflation \cite{Starobinsky, Whitt84}. In this regime, the predictions for the spectral index and tensor-to-scalar ratio are independent of $\xi$ and $V_J$ and take the universal values (up to subleading terms)
 \begin{align} \label{universal}
  n_s = 1 - 2/N_e + \ldots \,, \quad r = 12/N_e^2 + \ldots \,,
 \end{align}
which are in the sweet spot of the Planck results. The ratio $(\lambda/\xi)^2$ sets the amplitude of the CMB power spectrum. \\
\\
\textbf{Corrections.} As already emphasized in \cite{1310.3950}, it is natural to expect corrections to the attractor relation \eqref{1}. We will analyse the effect of such corrections on the Einstein frame potential and the attractor behaviour. To this end, we may consider either a correction in $\phi$ with the cost of having to specify an invertible $f(\phi)$ in order to obtain a relation $\phi(\chi)$, or - to allow for arbitrary $f(\phi)$ - simply parametrise a correction in terms of $f(\phi)$. Whereas the former leaves the coefficients of the series parametrizing the corrections unspecified, the latter maintains the beauty of the universal attractor, as any function $f(\phi)$ is allowed (examples are chaotic, natural and induced inflation, see \cite{1310.3950, Giudice14}; moreover, we have verified for the chaotic case that corrections in $\phi$ or $f(\phi)$ yield similar findings). Hence we consider the latter case and replace the attractor relation \eqref{1} with
\begin{align}
V(\phi)&=\lambda^2 h(f(\phi))^2 \label{h} \,,
\end{align}
where $\lambda^2$ remains a free parameter. The deviation of $h(f)$ from a linear function encapsulates the correction to the attractor relation. 
\begin{figure}[t!]
\centering
\includegraphics[scale=0.74]{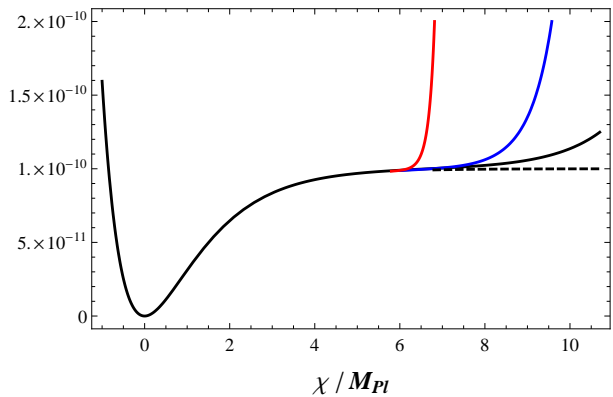}
\caption[V(Phi)]{\emph{Potential \eqref{16} with $n=1,2,8$ (black, blue, red from right to left). Higher $n$ require lower $\xi$ if one seeks just $55$ flat e-folds. The dashed line depicts the potential without corrections.}}
\label{potential}
\end{figure}

To develop a first intuition for such corrections, we consider a toy model with a single additional term. Taking
\begin{equation}
h(f)= f ( 1+c_n f^n) \,,
\end{equation}
and choosing $c_n\sim\mathcal O(1)$ in Planck units, one obtains 
\begin{equation}\label{16}
V_E=V_0\left[1+\mathcal O(1)\thinspace\xi^{-n}\left(e^{\sqrt{\frac{2}{3}}\chi} - 1\right)^n \right]^2,
\end{equation}
with $V_0$ being the unperturbed potential as in \eqref{staro}. Expanding the correction to leading order shows that its main contribution to the potential comes from a term $(\Omega/\xi)^n$. Hence the potential starts to deviate significantly from its plateau when the ratio $\Omega/\xi$ is greater than unity (as illustrated in Fig.~1). The point at which the deviation occurs is set through requiring the inflaton $\chi$ to traverse a certain distance in field space \emph{on the plateau} and enters the ratio through $\Omega$. In other words, a minimal length of a nearly flat plateau, or equivalently a required number of flat e-folds, translates into a lower bound of the coupling $\xi$ that is independent of $n$ for larger $n$. For lower values of $n$, $(\Omega/\xi)^n$ starts to contribute earlier than for higher $n$, hence $\xi$ increases in order to ensure $n_s(55)<0.980$. Thus any correction of order $n$ affects the attractor around the same point in field space for a value of $\xi$ that is set as to allow for at least $|N_e|$ flat e-folds. Corrections of higher power steepen the potential in a sharper way and thus the running of $n_s$ increases. Hence we find a larger running of $n_s$ to come from dominating higher-order terms in the correction. 

In order to quantify the above considerations, we have calculated the percentage of power suppression $\%(N_e)$ of $\Delta_s^2(k)$ at the onset of observable e-folds for exemplary values of $n$ (see Table I). In all cases, we have tuned $\xi$ such that $n_s(55)=0.970$; this is slightly higher than the universal value \eqref{universal} and hence signals the onset of the pre-Starobinsky phase of the scalar potential. Repeating this analysis for a redder or bluer $n_s(55)$ somewhat increases or decreases the non-minimal coupling $\xi$ respectively. \\
\\
\textbf{Examples.}
Understanding \eqref{h} not as a full UV theory but as an effective description, we consider not just a monomial correction but a series
\begin{equation}\label{series}
h(f)=\sum_{n=1}^{N}c_n f^n,
\end{equation}
where we again naturally assume all $c_n\sim\mathcal O(1)$. We find that a required amount of e-folds no longer translates into a lower bound on the coupling $\xi$. Instead, the coupling diverges linearly with the cutoff $N$. Furthermore, lower order terms dominate the steepening of the potential in the vicinity of the 55 e-folds point and hence the running of $n_s$ is weak regardless of any higher power terms in the series. Assuming a natural variation $\Delta c_n\sim \mathcal O(1)$ of the coefficients and thereby suppressing the first three terms of the series leads to an exemplary power suppression of about $2.5$\% to $4.8$\%, given a cut-off $N$=\thinspace $20$ and a coupling $\xi\sim\mathcal O(10^3)$ as to allow for at least 55 flat e-folds with $n_s(55)=0.970$. In this scenario, $N_{total}=115$. Thus understanding power suppression as a tool of effective field theory spectroscopy, we argue that a higher suppression indicates a cancellation or suppression mechanism of lower-order terms in the correction. 
\begin{table}[t]
\begin{tabular}{l*{7}{c}r}
$n$ & $\%(60)$ & $\%(62)$ & $\xi$ & $n_s(62)$ & $N_{total}$\\
\hline
$1$   	& $1.9$ & $3.4$ & $\mathcal O(10^{4})$ & $0.975$ & $272$ \\
$2$   	& $2.0$ & $3.8$ & $\mathcal O(10^{3})$ & $0.976$ & $173$  \\
$8$   	& $3.7$ & $7.7$ & $\mathcal O(10^{2})$ & $0.981$ & $90$\\
\end{tabular}
\caption{\emph{The effect of single higher-order corrections to the attractor relation. The suppression increases with $n$. Variation of $\xi$ compensates for the varying sharpness of the steepening. For larger $n$, $\xi\to\mathcal O(10^2)$ and $N_{total}$ approaches $\sim 62$.}}
\end{table}

To cure the divergent behaviour of $\xi$, we now seek to impose natural summation schemes in the expansion such that higher $c_n$ are effectively suppressed. We take
\begin{equation}\label{genericpert}
h(f)=  f \sum_{n=0}^{N}\frac{c_n}{n!} f^n,
\end{equation}
which may be understood as requiring that higher order terms are of decreasing importance. Without a cut-off and taking all $c_n =1$, the above yields
\begin{equation}\label{niceresult}
V_E=V_0 e^{\frac{2}{\xi}(\Omega-1)} \,.
\end{equation}
Again the requirement of a minimum number of flat e-folds translates to a lower bound on the coupling $\xi$. Allowing for at least $55$ flat e-folds and requiring $n_s(55)= 0.970$ induces a running of $n_s$ such that the power suppression is about 2.0\% to 3.6\%. More importantly, the above translates into a value of the coupling $\xi\sim\mathcal O(10^5)$, which is, having a natural $\lambda^2\sim\mathcal O(1)$, the required value to fit the normalisation of the power spectrum (see e.g.\ \cite{1402.1476}). Hence we find the normalisation of the power spectrum as well as the level of power suppression to be linked to the parameter $\xi$, which in turn is set by the amount of e-folds we require before any significant deviation from the nearly flat plateau occurs. Here, $N_{total}=264$. Truncating \eqref{genericpert} after the first 10 terms yields the same results, hence we conclude that higher order terms are phenomenologically negligible. In fact, provided the first few $c_n$ are of order one and given some cutoff, higher-order coefficients may be completely arbitrary. 

To study more exemplary corrections, consider a $\mathbb Z_2$ symmetry, i.e.\ we only invoke even terms in the correction. Applying this to \eqref{genericpert} yields $h = f \cosh(f)$ and
\begin{equation}\label{cosh}
V_E=V_0\cosh^2\left[ \xi^{-1}(\Omega-1)\right].
\end{equation}
In this case, tuning $\xi$ such that $n_s(55)=0.970$ gives a suppression of about 2.2\% to 4.0\%, where $\xi\sim\mathcal O(10^3)$. Considering a natural variation $\Delta c_n\sim\mathcal O(1)$ such that the first few lower order terms are suppressed and mimicking this by omitting the first two terms in the series expansion of the hyperbolic cosine, we find the suppression level to be increased to 3.1\% to 6.1\% where $\xi\sim\mathcal O(10^2)$. Hence scenarios with stronger suppression due to omitted lower order terms in the correction yield a sufficient amount of flat inflationary e-folds already for $\xi<\mathcal O(10^5)$. 

Finally, we vary our ansatz for \eqref{h} and consider the Jordan frame potential as a power series in $f(\phi)$, i.e.\
\begin{equation}\label{22}
h(f)=\sum_{n=1}^N \frac{c_n}{n!} f^n \,.
\end{equation}
The natural example of $h = e^f -1$ gives a suppression of up to 3.0\% and $n_s(55)=0.970$ for $\xi\sim\mathcal O(10^4)$. Restricting to only odd terms we have $h = \sinh(f)$ and find $n_s(55)=0.970$ with a suppression of up to 4.4\% for $\xi\sim\mathcal O(10^3)$. Considering the first five terms of a sum as in \eqref{22} without suppressing coefficients gives a power loss of up to 2.9\% and $n_s(55)=0.970$ for $\xi\sim\mathcal O(10^4)$. 

Remarkably,  in all examples considered in this section, the condition of 55 flat e-foldings translates into a range for the non-minimal coupling of the order $10^3$ up to $10^5$, depending on the specific correction. The upper end of this range leads to a power spectrum amplitude in concordance with the measured value. In contrast, the lower end of this range would have a larger amplitude. However, these values were obtained by requiring exactly 55 flat e-foldings and no more; tuning $\xi$ to the observed value around $10^5$ would simply lead to a longer inflationary plateau in these cases, and hence more flat e-foldings. Thus, the requirement of {\it at least} 55 flat e-foldings is remarkably consistent with the observed amplitude of the power spectrum, for a range of different examples: none of these examples require a non-minimal coupling of the order $10^6$ or higher to have a sufficiently long inflationary plateau for 55 e-foldings. \\
\\
\textbf{Higgs  Inflation.}
Our results also apply to Higgs inflation \cite{Salopek, Bezrukov08}.  This model has a non-minimal coupling set by $f = \phi^2$ and a scalar potential that takes the form, including corrections of type \eqref{genericpert},
 \begin{align}
 V_J=\lambda^2\left[\phi^4+ c_6 \phi^6+ c_8 \phi^8+\mathcal O(\phi^{10}) \right] \,.
 \end{align}
Taking both $\lambda^2$ and the coefficients $c_i$ of order one, there is a relation between the power normalisation and $N_e$. As shown in the previous section, both point towards a large non-minimal coupling (up to $10^5$). Claims that such a large coupling leads to a cutoff scale $M_p / \xi$ \cite{Burgess, Barbon} that is problematically close to or lower than the 
inflationary scale have been addressed in various ways \cite{Ferrara, Bezrukov11, Giudice14, Prokopec}. The vacuum stability regarding higher-order terms was discussed in \cite{Branchina}.
A different, logarithmic correction was considered in the
context of Higgs inflation at the critical point \cite{Bezrukov14,Hamada14}. While they also rely
on this correction to perturb the inflationary plateau, their aim and results are different: the
non-minimal coupling is chosen such that the correction affects the
entire observable period of the inflationary regime. Accordingly, they
require a lower $\xi$ and end up with predictions different from
\eqref{universal}. \\ 
\\ 
\textbf{Eternal Inflation.} If quantum fluctuations of the inflation field $\delta\phi_Q=H/(2\pi)$ dominate over the classical variation $\delta\phi_C=\dot\phi/H$, on average half of the fluctuations drive the field upwards its potential. Thus if a potential supports a regime where $H^2/({2\pi\dot\phi})>1$, inflation globally never ends \cite{0702178}. It can be shown that potentials have to support at least 1000 e-folds of inflation for the above scenario to be realised \cite{1404.4614}.  As demonstrated, natural corrections to the Universal Attractor yield models, where generically $N_{total}<300$, hence quantum fluctuations will always remain sub-dominant to the classical evolution. Thus if no high energy effects restore the flatness of the potential at large $\chi$, slow-roll eternal inflation appears disfavoured in this scenario. This is similar to the landscape \cite{Freivogel}, where $N_{total}$ is generally not much larger than the minimal amount of e-folds required.
\\ \\
\textbf{Discussion.} In this paper we studied the effects of corrections to a universal class of inflation models. Remarkably, these provide a link between the observed normalisation of the power spectrum and the number of flat e-foldings: both the height and length of the inflationary plateau are determined by the non-minimal coupling parameter $\xi$, which is required to be around or below $10^5$. We stress that, given either some cut-off or suppression mechanism, this single parameter determines the spectral index and amplitude of the power spectrum, the tensor-to-scalar ratio as well as the number of flat e-foldings. Moreover, for a range of corrections we predict a power loss of a few percent at low-$\ell$ in the temperature power spectrum of the CMB. It will be very exciting to put this expectation to the test in upcoming observational results, e.g.~the second-year Planck release.\\
\\
\small \textbf{Acknowledgements.} BB and AW are supported by the Impuls und Vernetzungsfond of the Helmholtz Association of German Research Centres under Grant No. HZ-NG-603. BB would like to thank the University of Groningen theory group for their hospitality during his stay.

\bibliography{prl-draft_17}

\end{document}